# Ethics, Data Science, and Health and Human Services: Embedded Bias in Policy Approaches to Teen Pregnancy Prevention

**Davon Woodard**
Ph.D. Student
Global Forum on Urban and Regional Resilience
PGG and Urban Computing
Virginia Tech
Blacksburg, VA

**Huthaifa I. Ashqar** (Corresponding author)
Booz Allen Hamilton
Washington, DC
hiashqar@vt.edu

**Taoran Ji**
Ph.D. Student
Department of Computer Science
Virginia Tech
Blacksburg, VA

# Abstract

**Background:** This study aims to evaluate the Chicago Teen Pregnancy Prevention Initiative delivery optimization outcomes given policy-neutral and policy-focused approaches to deliver this program to at-risk teens across the City of Chicago.

**Methods:** We collect and compile several datasets from public sources including: Chicago Department of Public Health clinic locations, two public health statistics datasets, census data of Chicago, list of Chicago public high schools, and their Locations. Our policy-neutral approach will consist of an equal distribution of funds and resources to schools and centers, regardless of past trends and outcomes. The policy-focused approaches will evaluate two models: first, a funding model based on prediction models from historical data; and second, a funding model based on economic and social outcomes for communities.

**Results:** Results of this study confirms our initial hypothesis, that even though the models are optimized from a machine learning perspective, there is still possible that the models will produce wildly different results in the real-world application.

**Conclusions:** When ethics and ethical considerations are extended beyond algorithmic optimization to encompass output and societal optimization, the foundation and philosophical grounding of the decision-making process become even more critical in the knowledge discovery process.

# 1 Introduction

The City of Chicago is a leader in the urban data science and open data movements, being one of only a handful of municipalities in the US with a fully-staffed team addressing complex urban issues with data science, advanced analytics, and machine learning. Assuming the role of an external consultants to the DOIT, the Chicago Department of Public Health (CDPH), and Chicago Public Schools (CPS), the goal of our study is to provide compendium of the ethical issues faced in HHS, and to explore the impacts of one ethical consideration in the application of data science and analytics on the delivery services. In 2010, the Chicago Teen Pregnancy Prevention Initiative (TPPI), was launched with the aim "to reduce teen pregnancies and improve access to care for adolescents." The TPPI was a 19.7 million dollars funded program by the U.S. Department of Health and Human Services' Office of Adolescent Health (OAH), jointly administered program through the Chicago Department of Public Health (CDPH) and Chicago Public Schools (CPS).[1]

In general, TPPI has three levels of actions: city, schools, and students' level. In the city level, an action plan was created by health, policy and education experts who serve on the Adolescent Health Access Committee. The plan details more than 40 health measures and goals specific to adolescents. The action plan was distributed to principals, partners, and delegate agencies. It contains public awareness campaigns and creating a data repository for adolescent health. In the school level, CPS Board adopted new policy that mandates comprehensive sexual health education for grades K through 12. This comprehensive curriculum should be age appropriate and aligned with the National Sexuality Education Standards. In the student level, a plan was created with the corporation of Teen Outreach Program (TOP) and Youth Advisory Council. TOP is an evidence-based positive youth development program, shown to reduce teen pregnancy, course failure, and suspensions by half. The TOP curriculum includes lessons on clarifying values and healthy relationships, communication/assertiveness, goal-setting and decision-making, human development and sexuality community service learning. The curriculum is taught by facilitators and certified by Wyman, the creator of the Teen Outreach Program (TOP).[1]

Stepping back to 2010, at the outset of the TPPI, our goal is to evaluate program delivery optimization outcomes given policy-neutral and policy-focused approaches to deliver this program to at-risk teens across the City of Chicago. The policy-neutral approach will consist of an equal distribution of funds and resources to schools and centers, regardless of past trends and outcomes. The policy-focused approaches will evaluate two models: first, a funding model based on prediction models from historical data; and second, a funding model based on economic and social outcomes for communities.

Using the points raised by Tene and Polonetsky[2], this study will explore policy decisions of data and decision managers influence to the outcomes of analytics projects by positioning data science approaches within specific frameworks. In doing this, we hope to advocate for the increased reflexivity of managers throughout the data science process, understanding out their own positionality and subjectively can have real and significant impacts on the lives of citizens.

## 2   Data Collection and Problem Setup

The City of Chicago's TPPI program was a part of a national Teen Pregnancy and Prevention initiative which provided funding to 75 state or local health departments and organizations.[3,4] The initiative evaluated ten programs, not including the Chicago TPPI. These ten programs served between 55 and 14,492 youth and provided services ranging in average cost between $68 and $11,541 per student. Unlike the reported programs, the TPPI in the City of Chicago was unique in that it did not provide intensive small-group prevention services, but instead provided city-wide education and prevention services targeted at both male and female students - as such, the average cost per student was calculated to be $38. This estimate was calculated based on the assumption of equal funding allocation per grant funded year and averaging that over all of 101,008 high school aged students in the first year of program funding. We collect and compile several datasets from public sources including American FactFinder[5] and City of Chicago Open Data[6]:

1. Chicago Department of Public Health Clinic Locations[7]: City of Chicago Mental Health, Sexually Transmitted Infection (STI) Specialty, and Women Infant Children (WIC) clinic locations, hours of operation and contact information.
2. Public Health Statistics I[8]: This dataset contains the annual number of births to mothers aged 15-19 years old and annual birth rate, by Chicago community area, for the years 1999-2009.
3. Public Health Statistics II: This dataset contains the annual number of births to mothers aged 15-19 years old and annual birth rate, by Chicago community area, for the years 2010-2014.
4. Census Data of Chicago[9]: This dataset contains general population and housing characteristics (e.g., population, age, sex, race, households and housing) of Chicago area.
5. Chicago Public High School Locations[10]: This dataset contains the location of all public high schools within the city of Chicago.

6. List of Chicago Public High Schools: This list of Chicago public high schools contains student number information of 130 schools which were active during the whole TPPI grant period, and their grade/total attendance for each year. There is a total of 130 schools. We take the total TPPI grant and divide equally over the years we get 3.94 million per year.

In this study, we mainly used public health statistics and list of Chicago public high schools, in which statistics of each year consists of four columns — teen births, teen birth rate, teen birth rate lower $CI$ and teen birth rate upper $CI$. We adopt the teen birth rate column and for each community area code $c \in \{1, \ldots, 76\}$, a time series of data $x = \{x_l, \ldots, x_{l+\Delta}\}$ is generated. In our collected data, $l = 1999$ and $\Delta = 14$. Also, we get another data flow used to estimate the TPPI grant spent at each school each year. After preprocessing, the students' number in each school is mapped to the number of teenagers of each community and a time series data $y = \{y_l, \ldots, y_{l+\Delta}\}$ is generated. Thus, the goal of our study is presented as predict the teenage pregnancy rate $x_r$ given the above two time series $x$ and $y$ which are either directly or indirectly correlated with the teenage pregnancy rate of each community.

## 3 Methods

In this study, we will focus on the exploration of usage of the autoregressive model. Dataset collected in this study are mostly presented in the format of time series, which inspires us to consider the use of autoregressive (AR) model under the assumption that output variable and its previous values are linearly dependent as shown in (1).

$$x_t = \sum_{i=1}^{l} w_i x_{t-1} + \epsilon \qquad (1)$$

where $w_i, i = 1, \ldots, l$ is the parameter controlling the influence of "momentum" value over current value and $\epsilon$ is white noise. AR model has several variants including ARMA[11] (autoregressive–moving-average model) which combines both autoregression and moving average, and ARX[12] (autoregressive exogenous) model which consider both data's previous values and related exogenous data flows. Apart from the historical data of teenager birth rate from 1999 to 2009 in different communities in Chicago, we have the TPPI spending strategy plan for different schools in Chicago. Taking into account that in the real-world scenario TPPI grant is supposed to be spent to reduce the teenager pregnancy rate, then by assuming a negative linear correlation between TPPI grant money and teenager pregnancy rate, we combine the data flow of historical TPPI grant spending strategy into the existent AR model and generate a series of data which fit the definition of ARX model as shown in (2).

$$x_t + w_1 x_{t-1} + \cdots + w_l x_{t-l} = b_1 y_{t-u} + \cdots + b_v y_{t-u-v}$$
$$\Rightarrow x_t = -w_1 x_{t-1} - \cdots - w_l x_{t-l} + b_1 y_{t-u} + \cdots + b_v y_{t-u-v} \qquad (2)$$

where $x = \{x_1, \ldots, x_i, \ldots\}$ is the time series of historical teen birth rate, and $y = \{y_1, \ldots, y_i, \ldots\}$ is the time series of TPPI grant every year.

## 4  Evaluation and Discussion

Based on how many historical records are considered, i.e., the value of $l, u$, and $v$, we construct and train different autoregressive models. In the first experiment, AR model is applied to the time series of teenager birth rate to observe if different data points are linear dependent. In the second experiment, ARX model is applied to analyze the relatedness between TPPI grant allocation and teenager birth rate.

## 4.1 Relatedness between Historical Teenager Birth Rate Data

In this section, we explore the relatedness among time series of historical teen birth rate to understand if there does exist a linear relation among them as shown in the AR model. Meanwhile, it's helpful for us to understand that TPPI grant worked as expected, that is, reducing teen birth rate. Under this experiment settings, we train five models with $l \in \{1, 2, 3, 4, 5\}$. First, overall correlation coefficient ($R^2$) score is evaluated for five models and is shown in Figure 1. $R^2$ value steadily increases when models are fed more data from historical years. And there exists an exactly linear fit in time series of teen birth rates when $l$ reaches five. That means, given five years historical data of teen birth rates, an autoregressive model can almost perfectly predict the teen birth rate in a certain area. Secondly, we compute the $R^2$ values for different models for different areas and list the top ten best-predicted areas and corresponding $R^2$ scores in Table 1. From the result, for some community areas such as Edgewater, Douglas, AR models perform well even when pretty few years of historical data are considered.

Based on the above results, we obtain a conclusion that for a certain community area, linear relatedness exists among the time series of teen birth rates. Also, considering that the model does the prediction without considering the actual TPPI grant allocation and still obtains a stable linear combination. This shows that either TPPI grant allocation is stable distributed for each year or it is not very effective in influencing the teen birth rate.

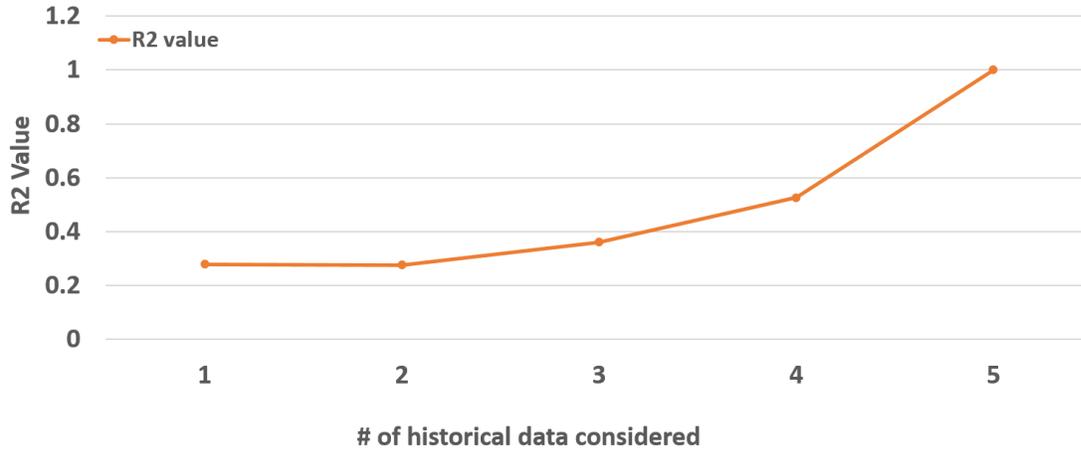

Figure 1. Overall $R^2$ performances of models considering different years of historical data.

Table 1. Top ten close-to-fact predictions of each model.

| Index | 1-year historical data | | 2-year historical data | | 3-year historical data | | 4-year historical data | |
|---|---|---|---|---|---|---|---|---|
| 1 | Edgewater | 0.77 | Douglas | 0.81 | Near South Side | 0.88 | Norwood Park | 0.97 |
| 2 | Grand Boulevard | 0.75 | North Center | 0.78 | West Pullman | 0.86 | Garfield Ridge | 0.97 |
| 3 | Lake View | 0.74 | Edgewater | 0.69 | North Center | 0.84 | South Lawndale | 0.94 |
| 4 | West Town | 0.71 | Grand Boulevard | 0.66 | Douglas | 0.82 | Lincoln Square | 0.93 |
| 5 | Rogers Park | 0.68 | Belmont Cragin | 0.66 | Hyde Park | 0.77 | Near South Side | 0.92 |
| 6 | Douglas | 0.67 | Albany Park | 0.64 | Edgewater | 0.77 | North Park | 0.87 |
| 7 | Riverdale | 0.65 | Lake View | 0.62 | Garfield Ridge | 0.70 | West Lawn | 0.87 |
| 8 | Bridgeport | 0.65 | Riverdale | 0.61 | Roseland | 0.64 | West Pullman | 0.86 |
| 9 | North Center | 0.63 | Near North Side | 0.57 | North Park | 0.63 | Douglas | 0.85 |
| 10 | Washington Park | 0.63 | Armour Square | 0.56 | Belmont Cragin | 0.62 | Belmont Cragin | 0.85 |

## 4.2 Relatedness between Teenage Birth Rate and TPPI Grant Allocation

In this section, we explore the relatedness between two time series: history data of teen birth rate and historical data of TPPI grant allocation records, and plan to understand if they are linearly dependent as shown in the ARX model. Specifically, we fix $u$ to 1, that is, using only one historical data. Also, we choose only overlapped years between two time series, which is from 2010 to 2014, as data ingest. Then four models are trained considering different historical years, i.e., $\langle l = 1, v = 1 \rangle, \langle l = 2, v = 1 \rangle, \langle l = 2, v = 2 \rangle$ and $\langle l = 1, v = 2 \rangle$. Overall $R^2$ score is evaluated for the four models as shown in Figure 2. When either of two parameters ($l$ and $v$) is larger than one, these two time series will almost perfect fit to a linear equation, which implies hidden linear dependences.

In this case, it is important to understand how TPPI grant contribute to the control of teenage birth rate for each community area. Table 2 shows how the teen birth rate is influenced by the last year TPPI grant allocation. For example, every one thousand dollars spent in Greater Grand Crossing will help to decrease 5.9 in teen birth rate for next year. Results also shows that Greater Grand Crossing are in top 10 in all the four models, and the rank value for Irving Park is very close to zero in most cases. This proves that TPPI fund might not be equally effective to all community areas.

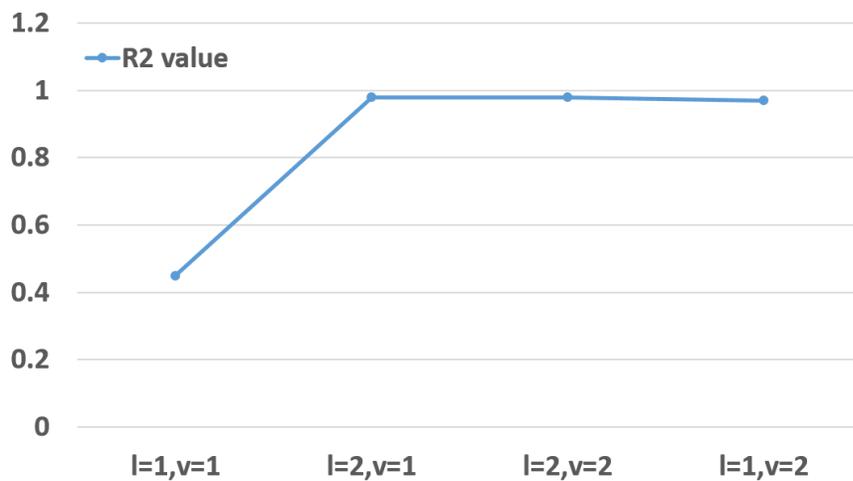

Figure 2. Overall $R^2$ performances of models considering different years of historical data.

Table 2. Top 10 communities which TPPI grant are most effective.

| Index | $l = 1, v = 1$ | | $l = 2, v = 1$ | | $l = 1, v = 2$ | | $l = 2, v = 2$ | |
|---|---|---|---|---|---|---|---|---|
| 1 | Rogers Park | -0.93 | Greater Grand Crossing | -5.90 | Grand Boulevard | -7.80 | Greater Grand Crossing | -2.00 |
| 2 | Greater Grand Crossing | -0.91 | Douglas | -1.12 | Washington Park | -5.99 | Grand Boulevard | -0.72 |
| 3 | Douglas | -0.51 | Grand Boulevard | -0.51 | Englewood | -3.79 | Garfield Ridge | -0.48 |
| 4 | East Side | -0.48 | West Ridge | -0.27 | Greater Grand Crossing | -2.40 | Rogers Park | -0.45 |
| 5 | Edgewater | -0.18 | Edgewater | -0.26 | Irving Park | -2.03 | Douglas | -0.35 |

| 6 | West Ridge | -0.16 | Garfield Ridge | -0.14 | Garfield Ridge | -1.01 | Woodlawn | -0.33 |
|---|---|---|---|---|---|---|---|---|
| 7 | West Town | -0.13 | Rogers Park | -0.06 | Morgan Park | -0.72 | Near West Side | -0.24 |
| 8 | Archer Heights | -0.09 | Near West Side | -0.04 | Rogers Park | -0.55 | Archer Heights | -0.18 |
| 9 | Humboldt Park | -0.08 | Irving Park | -0.04 | Woodlawn | -0.50 | West Ridge | -0.14 |
| 10 | Grand Boulevard | -0.04 | West Town | -0.04 | Near West Side | -0.30 | Irving Park | -0.13 |

# 5 Conclusion

Results of this study confirms our initial hypothesis, that even though the models are optimized from a machine learning perspective, there is still possible that the models will produce wildly different results in the real-world application. In the application of data science to urban contexts, specifically health and human service delivery, given the expansive impact that these services, or denial of service can have, it is important for policymakers and programmers to be reflexive in regard to structure not only the algorithm, but the philosophical foundations underpinning the algorithms, as those too have significant impact on social well-being.

When ethics and ethical considerations are extended beyond algorithmic optimization to encompass output and societal optimization, the foundation and philosophical grounding of the decision-making process become even more critical in the knowledge discovery process. Given the scope and long-term intergenerational impact of health and human services at the city-wide level, the aim of this project was to evaluate how changes in the initial stages of the knowledge discovery process, creating policy-neutral and policy-focused algorithms effect the social outcomes. Using a series of autoregressive (AR) models on geolocated population and demographic data from the US Census and Chicago Public Schools layered with historical City of Chicago Department of Health and Human Services teen pregnancy data the results demonstrated that even with nearly perfect algorithmic optimization, the social impacts and outcomes can vary widely.

## Acknowledgment

This work is supported in part by the National Science Foundation via grant #DGE-1545362, UrbComp (Urban Computing): Data Science for Modeling, Understanding, and Advancing Urban Populations.

## Statement on conflicts of interest

The authors declare that there is no conflict of interest.

# 6 Summary

1. This study aims to evaluate the Chicago Teen Pregnancy Prevention Initiative delivery optimization outcomes given policy-neutral and policy-focused approaches to deliver this program to at-risk teens across the City of Chicago.

2. Results of this study confirms there is still possible that Machine Learning models will produce wildly different results in the real-world application.

3. Positionality and subjectively in data-driven health policies might have real and significant impacts on the lives of citizens.